\documentclass[notitlepage,aps,prl,reprint,twocolumn,longbibliography,superscriptaddress]
{revtex4-2}
\usepackage{graphicx}
\usepackage{amsmath}
\usepackage{amssymb}
\usepackage{comment}
\usepackage[colorlinks, allcolors=blue]{hyperref}
\usepackage[all]{hypcap}
\usepackage[mathlines]{lineno}
\usepackage{physics}
\usepackage{wrapfig}
\usepackage{lipsum}
\usepackage[normalem]{ulem}
\usepackage{siunitx}

\newcommand{\pref}[2]{\hyperref[#1]{\ref{#1}(#2)}}

\begin{document}

\title{Programmable Gauge-Field Textures with Ultracold Atoms in Momentum Space}

\author{Hongru Wang}
\thanks{These authors contributed equally to this work}
\affiliation{%
Zhejiang Key Laboratory of Micro-nano Quantum Chips and Quantum Control, School of Physics, and State Key Laboratory for Extreme Photonics and Instrumentation, Zhejiang University, Hangzhou, China
}%
\author{Hang Li}
\thanks{These authors contributed equally to this work}
\affiliation{%
Zhejiang Key Laboratory of Micro-nano Quantum Chips and Quantum Control, School of Physics, and State Key Laboratory for Extreme Photonics and Instrumentation, Zhejiang University, Hangzhou, China
}%
\affiliation{%
State Key Laboratory of Precision Spectroscopy, Institute of Quantum Science and Precision Measurement, East China Normal University, Shanghai 200062, China
}%
\author{Yichen Pan}
\author{Yuyan Luo}
\affiliation{%
Zhejiang Key Laboratory of Micro-nano Quantum Chips and Quantum Control, School of Physics, and State Key Laboratory for Extreme Photonics and Instrumentation, Zhejiang University, Hangzhou, China
}%
\author{Bryce Gadway}
\affiliation{Department of Physics, The Pennsylvania State University, University Park, Pennsylvania 16802, USA}
\author{Tao Chen}
\email{ taochen@xjtu.edu.cn}
\affiliation{School of Physics, Xi’an Jiaotong University, Xi’an 710049, China}
\author{Bo Yan}
\email{yanbohang@zju.edu.cn}
\affiliation{%
Zhejiang Key Laboratory of Micro-nano Quantum Chips and Quantum Control, School of Physics, and State Key Laboratory for Extreme Photonics and Instrumentation, Zhejiang University, Hangzhou, China
}%
\date{\today}

\begin{abstract}

Synthetic gauge fields with ultracold atoms offer a route to quantum matter in which electromagnetic environments can be designed rather than merely imposed. While the Harper–Hofstadter model has been realized in several cold-atom systems, existing implementations are largely limited to spatially uniform magnetic fluxes. Here we experimentally realize a highly programmable two-dimensional momentum-state lattice of ultracold atoms with local control over the Peierls phase pattern, enabling direct implementation of Harper–Hofstadter Hamiltonians with tunable and spatially structured synthetic gauge fields. We observe a crossover from ballistic to strongly flux-modified bulk dynamics with suppressed transport. By introducing a synthetic electric field through site-dependent energy gradients, we further demonstrate Hall-type transverse drift arising from the interplay between electric and magnetic fields. In addition, we engineer a synthetic flux domain wall separating regions with opposite magnetic fluxes and observe anisotropic propagation guided along the interface. These results move cold-atom gauge-field engineering from uniform magnetic backgrounds toward designer gauge textures, providing an experimental setting for transport across programmable topological interfaces.
\end{abstract}
\maketitle

Responses of charged particles under electromagnetic fields are crucial for understanding a broad range of fundamental phenomena, from quantum Hall effect to transport behaviors under frustration in topological quantum materials \cite{2010_RMPTI1,2011_RMPTI2,Haldane2017,Ozawa2019,Haldane1988,Vidal1998,Struck2011,Jotzu2014,Leykam2018}. Over the past decades, ultracold atoms have emerged as a versatile platform for engineering synthetic electromagnetic field to simulate these topics in highly controllable settings \cite{Cooper2008,2012_qugas,Goldman2016}. To date, two primary methods have been developed to imprint tunneling phases and generate synthetic magnetic fluxes for neutral atoms. 
One relies on periodic modulation of optical lattices \cite{Lignier2007,Struck2012,Struck2013,Weitenberg2021}. By carefully tuning the shaking amplitude and frequency, one can effectively achieve tunable tunneling phases that allow for simulating the celebrated Harper-Hofstadter (HH) model \cite{Aidelsburger2013,Miyake2013}, e.g., explorations of the phase-driven bulk topology \cite{Aidelsburger2015,Kennedy2015} and fractional quantum Hall state under many-body interactions \cite{Tai2017,Leonard2023}. Another route is via synthetic dimension approaches \cite{Lin2009,Celi2014,Arguello2024,Yu2025,Ren2026}. For example, the Harper-Hofstadter lattices have also been realized in hybrid two-dimensional (2D) synthetic systems, with tunneling phase introduced along the internal hyperfine-state dimension using two-photon Raman transitions, which enables direct observations of chiral edge currents \cite{Stuhl2015,Mancini2015,Bouhiron2024} and further universal Hall responses in strongly correlated regime \cite{Zhou2023}. 

However, existing implementations are largely constrained to spatially uniform gauge fluxes because, for example, the Raman lasers are globally applied and lack site-resolved selectivity. Many phenomena of current interest, including transport across magnetic domain walls, spatial gauge textures and interfaces between distinct topological regions, require control of the flux pattern in real or synthetic space rather than control of only its global value. Extending gauge field control toward spatially programmable local and spatially inhomogeneous flux configurations therefore remains an important frontier. Such capability would enable investigations of quantum transport across engineered topological interfaces and open new possibilities for simulating complex synthetic quantum materials with designed gauge structures. Momentum-state lattices (MSLs) with ultracold ataoms have recently been shown as a powerful platform for programmable synthetic lattice engineering \cite{Gadway2015,Meier2016, 2018_science, 2019_npj}. While local gauge field control has been demonstrated in several quasi-one-dimensional geometries \cite{An2018zigzag,An2017chiral,Gou2020,li2022aharonov,Liang2022,Li2023,Liang2024,Zeng2024,li2025engineering}, the realization of genuinely 2D MSLs with controlled plaquette flux patterns is still a pressing goal \cite{Agrawal2024} but remains experimentally challenging and elusive. 

\begin{figure*}[]
	\centering
	\includegraphics[width=0.95\linewidth]{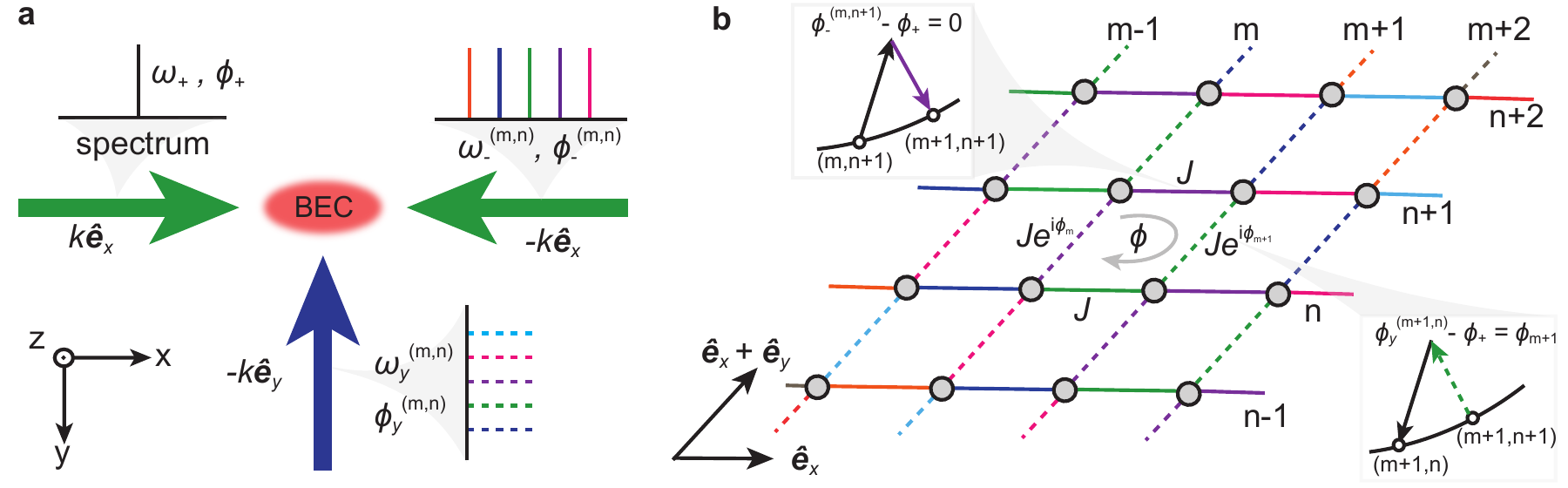}
    \caption{\textbf{Schematic of building 2D MSLs with tunable magnetic flux.} 
    \textbf{(a)} Three Bragg laser beams (the wavevector $k=2\pi/\lambda$ with $\lambda\approx 794.7~{\rm nm}$) interact with a BEC to form 2D lattice geometries in momentum space. The two counter-propagating beams along the $x$-direction ($\omega_+$ and $\omega_-$) trigger the tunnelings between discrete momentum states along $\hat{e}_x$ with an interval of $2\hbar k$, while the perpendicular beam $\omega_y$, in combination with $\omega_+$, leads to hopping between states with a momentum difference of $\hbar k(\hat{e}_x + \hat{e}_y)$. The tunneling phases are introduced by programming each individual phase for different frequency components in $\omega_-$ and $\omega_y$ beams, respectively. 
    \textbf{(b)} Illustration of the Haper-Hofstadter lattice. Bonds with the same color and the same line shape share the same hopping amplitude and phase, as they are triggered by the same two-photon Bragg process (see Supplementary Material for details on momentum and energy selection rules \cite{supp}). Here we simply let all hoppings along $\hat{e}_x$ direction (solid connections) take zero phase, i.e., $\phi_-^{(m,n)}-\phi_+ = 0$. For the center plaquette, the magnetic flux $\phi$ is determined by the tunneling phases of $(m,n)\to (m, n+1)$ and $(m+1, n+1)\to (m+1, n)$ transitions, i.e., $\phi=\phi_{m+1} - \phi_m$ where $\phi_m$ indicates the phase difference between the $\omega_y^{(m,n)}$ component and the $\omega_+$ beam. }
	\label{fig1}
\end{figure*}

Here we experimentally demonstrate highly programmable 2D MSLs using a Bose–Einstein condensate (BEC), and further implement Harper–Hofstadter model Hamiltonians with local control of the gauge flux on each individual plaquettes by imprinting Peierls phases. The central advance is that the magnetic flux is no longer fixed by a single global driving phase, but can be written as a programmable spatial pattern through the phases of selected Bragg-frequency components. We emphasize that the link-resolved phase programming requires more precise and coherent control over the two Bragg beams along different spatial directions, which is totally different from our previous implementation of 2D MSLs where Bragg beams in two directions do not coherently interact with each other and thus necessarily lead to zero net flux \cite{Dong2025}. Starting from the bulk response of a uniform-flux Harper–Hofstadter lattice, we further observe Hall-type transverse drift induced by an additional synthetic electric field produced by a site-dependent energy gradient. Moreover, we explicitly probe the anisotropic dynamical evolution behaviors along a phase-domain-wall boundary that separates two lattice regions with opposite magnetic flux. The combination of uniform-field benchmarking, Hall deflection and interface-guided dynamics establishes the same platform as both a simulator of Hofstadter transport and a tool for engineering gauge-field textures. Our work establishes a versatile route toward studying 2D programmable synthetic quantum matter and paves the way for engineering novel tailored topological environments beyond uniform gauge-field settings.

~

\begin{figure*}[]
	\centering
	\includegraphics[width=\linewidth]{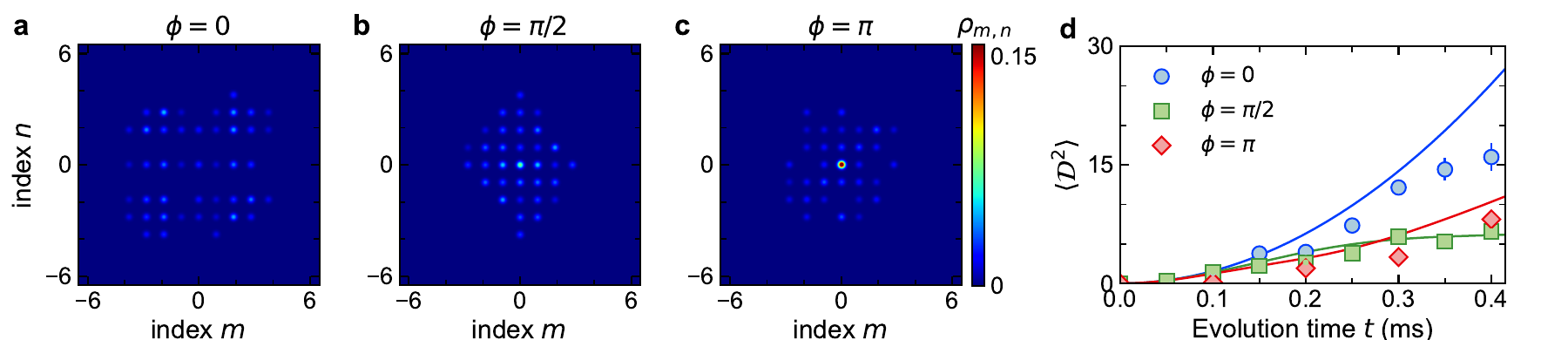}
    \caption{
    \textbf{Bulk dynamics for an injection in 2D Harper-Hofstadter lattices.}
    \textbf{(a-c)} Measured population distributions $\rho_{m,n}$ at $t=0.3~{\rm ms}$ for (a) $\phi=0$, (b) $\phi=\pi/2$ and (c) $\phi=\pi$. The wavepacket spreading shows nearly symmetric ballistic behavior in the absence of flux, while the spreading speed gets suppressed under finite $\phi$-flux.
    \textbf{(d)} Time evolution of the bulk expansion characterized by the mean-square displacement $\langle \mathcal{D}^2\rangle$ for different gauge fluxes $\phi$. For zero flux, the wavepacket exhibits ballistic expansion with $\langle\mathcal{D}^2\rangle\propto t^2$, whereas a slower expansion is observed for $\phi=\pi/2$ and $\phi=\pi$. Here the tunneling amplitudes are set as $J/h = 0.95(5)~{\rm kHz}$ for a $15\times 15$ square lattice. 
    All solid lines indicate the ideal Hamiltonian simulation results with experimental parameters, and the error bars are from multiple independently measured data sets.
    }
	\label{fig2}
\end{figure*}

\emph{Harper-Hofstadter model implementation. --} As shown in Fig.~\pref{fig1}{a}, we make use of three Bragg laser beams to induce momentum transfer to form 2D MSLs. Building upon our previous one-dimensional experimental setup \cite{2019_npj,Xie2020,Gou2020,Dong2024,2021_npj} where two counter-propagating beams interact with a $^{87}$Rb BEC containing $\sim 4\times 10^4$ atoms, here we introduce an additional beam propagating perpendicular to the original Bragg axis to connect discrete momentum states in a second synthetic direction. By considering the selection rules of both momentum and energy conservations for two-photon Bragg processes, the two beams with wavevectors $-k\hat{e}_x$ and $-k\hat{e}_y$ contain multiple frequency components. The component indexed by $(m,n)$, together with the $k\hat{e}_x$ beam, accounts for the $\ket{2m\hbar k \hat{e}_x} \to \ket{2(m+1)\hbar k\hat{e}_x}$ and $\ket{n\hbar k (\hat{e}_x + \hat{e}_y)} \to \ket{(n+1)\hbar k (\hat{e}_x + \hat{e}_y)}$ connections, respectively. The laser phases of the $(m,n)$ components, denoted as $\phi_-^{(m,n)}$ and $\phi_y^{(m,n)}$, and the $k\hat{e}_x$ beam, $\phi_+$, directly imprint effective Peierls phases for each corresponding tunneling link. In this way, a set of discrete momentum states $\ket{(2m+n)\hbar k\hat{e}_x + n\hbar k \hat{e}_y}$, uniquely labeled by two integers $(m,n)$, form an effective 2D lattice structure; see Fig.~\pref{fig1}{b}.

To implement the Harper-Hofstadter Hamiltonian, we calibrate all tunneling amplitudes to a uniform value of $J$, while retaining full control over the tunable tunneling phases. Here we choose $\phi_-^{(m,n)} = \phi_+$ for all links along the $\hat{e}_x$ direction, while the links along the $(\hat{e}_x+\hat{e}_y)$ direction satisfy $\phi_y^{m,n} - \phi_+ = \phi_{m}$, as illustrated in Fig.~\pref{fig1}{b}. Then, the accumulated magnetic flux in each plaquette is determined by $\phi=\phi_{m+1} - \phi_m$. The effective single-particle Hamiltonian reads
\begin{equation}
 \mathcal{H}_{\rm HH} = -J\sum_{m,n}\left(e^{i\phi_m}c_{m,n+1}^\dagger c_{m,n}^{} + c_{m+1,n}^\dagger c_{m,n}^{}\right)  + {\rm H.c.},
\end{equation}
where $c^\dagger_{m,n}$ indicates the creation operator at site $(m,n)$ with a momentum ${(2m+n)\hbar k\hat{e}_x + n\hbar k \hat{e}_y}$. To note, while we focus on both uniform tunneling amplitudes and magnetic fluxes here, the programmable capability of this three-beam scheme is considerably general. For example, we can realize a phase domain-wall boundary that separates two lattice regions with opposite magnetic fluxes; see Supplementary Material \cite{supp}. 

It is worth mentioning that our phase engineering scheme is still constrained to the degeneracy between different Bragg transitions. Since only $m+n$ independent laser components are used to control $m\times n$ tunneling links in 2D, each component actually responds to multiple transitions simultaneously \cite{supp}. For example, a given $(m,n)$ component in the perpendicular $y$-direction beam drives a family of two-photon processes $\ket{m-\ell,n+\ell} \to \ket{m-\ell,n+\ell+1}$ for $\ell\in\mathbb{Z}$. Accordingly, the present implementation should be viewed as programmable over a broad and experimentally useful class of flux textures, rather than as an unconstrained independent knob on every microscopic link. This distinction is important because the demonstrated uniform fields, electric-field response and opposite-flux domain wall all lie within this controllable class and are directly calibrated by the same phase-programming protocol. Nevertheless, the present scheme establishes a highly flexible and programmable platform for exploring Harper–Hofstadter model physics with ultracold atoms, where both lattice geometry and synthetic gauge structure can be engineered in a controlled manner. Similar to one-dimensional case, the time evolution of the population in each $(m,n)$ site can also be read out in parallel via time-of-flight measurements. 

~

\emph{Flux-dependent bulk dynamics. --} 
We first benchmark the implemented Harper-Hofstadter lattice by probing the responses of a localized bulk injection under different uniform magnetic fluxes. The atoms are initialized at the center $(m =0, n=0)$ site in a $15\times 15$ lattice with homogeneous tunneling amplitudes $J/h=0.95(5)~{\rm kHz}$. Since the gauge flux strongly modifies the underlying band structure and therefore leads to different effective spreading velocity, we anticipate to observe different transport behaviors when tuning the flux from zero to $\pi$. The effective spreading velocity directly relates to the mean-square displacement defined as
\begin{equation}
 \langle \mathcal{D}^2\rangle = \sum_{m,n}  (m^2 + n^2) \rho_{m,n},
\end{equation}
where $\rho_{m,n}$ is the population fraction on site $(m,n)$. Experimentally, we record the time evolutions of $\rho_{m,n}$, and then resolve the behavior of $\langle \mathcal{D}^2\rangle$ as a function of time to characterize the flux-dependent bulk dynamics. 

Figures \pref{fig2}{a-c} show the measured population distributions in each $(m,n)$ sites after an evolution time of $\sim 0.3~{\rm ms}$ for three representative flux values. For $\phi=0$, the wavepacket exhibits symmetric expansion along both lattice directions, consistent with isotropic transport in the square lattice geometry. The corresponding evolution of the mean-square displacement, shown in Fig.~\pref{fig2}{d}, displays a nearly quadratic scaling $\langle \mathcal{D}^2\rangle\propto t^2$, directly evidencing the ballistic free expansion. 
Finite gauge flux $\phi$ significantly modifies the bulk transport. As shown in Figs.~\pref{fig2}{b,c}, when $\phi=\pi/2$ and $\pi$, the population distributions become substantially more localized around the initial center site, with a pronounced suppression of the expansion velocity compared with the zero-flux case. This behavior reflects the emergence of cyclotron-like dynamics induced by the synthetic magnetic field, which constrains the propagation of the wavepacket through destructive quantum interference. Correspondingly, as shown in Fig.~\pref{fig2}{d}, the measured mean-square displacement varies approximately linearly over short evolution times, indicating a crossover from ballistic to diffusive regimes. The observed bulk response agrees well with theoretical predictions of the Harper–Hofstadter model, demonstrating the programmability of the synthetic gauge field in this 2D MSL platform.

~

\begin{figure*}[]
	\centering
	\includegraphics[width=0.9\linewidth]{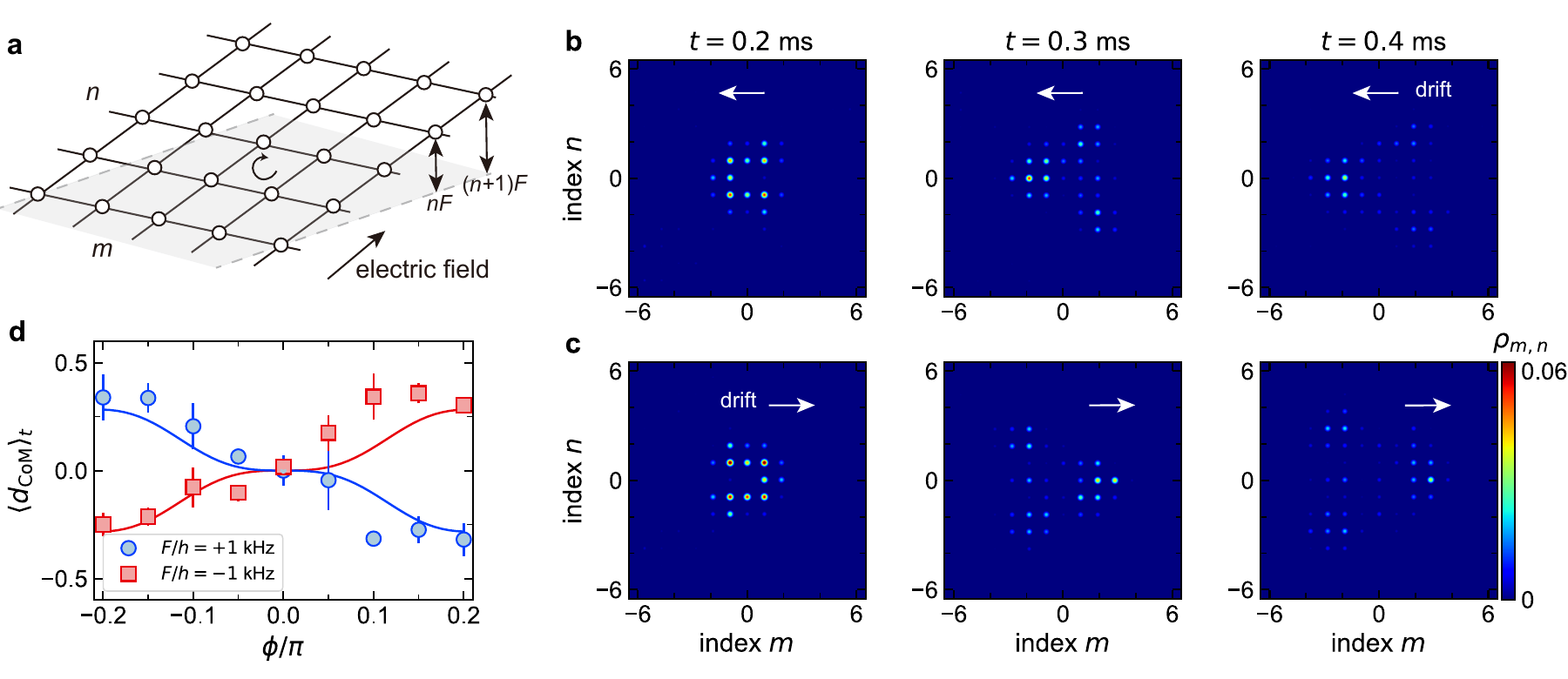}
	\caption{\textbf{Hall response under programmable transverse electric fields.} 
    \textbf{(a)} Schematic of simulating the Hall effect in the Harper-Hofstadter lattice. A potential gradient is applied along the $\hat{e}_x+\hat{e}_y$ direction (sites indexed by $n$) to introduce a synthetic electric field $F$.
    \textbf{(b,c)} Measured momentum-space dynamics of a wavepacket injected at the lattice center for evolution times $t=0.2, 0.3$ and $0.4~{\rm ms}$ with the effective electric fields $F/h=+1~{\rm kHz}$ and $F/h=-1~{\rm kHz}$, respectively, showing opposite transverse drift directions of the wavepacket. Here $\phi=\pi/5$ is used.
    \textbf{(d)} Time-averaged center-of-mass displacement $\langle{d}_{\rm CoM}\rangle_t$ versus the magnetic flux $\phi$ for the two opposite electric-field polarities, averaged over the evolution time of $t=0 ~{\rm ms}$ to $t=0.5~{\rm ms}$ \cite{supp}. Solid lines are numerical simulation results, and the error bars indicate the standard errors from multiple independently measured data sets.}
    \label{fig3}
\end{figure*}

\emph{Observation of Hall drift. --} 
A hallmark consequence of charged-particle dynamics in the Harper–Hofstadter model is the emergence of transverse Hall motion under an applied electric field, originating from the interplay between the external force and the underlying magnetic flux. Thanks to the flexible programmability of on-site potentials in MSLs, our platform enables direct observation of such Hall-type transport dynamics. Experimentally, we generate a synthetic electric field by introducing a linear potential gradient along the $(\hat{e}_x+\hat{e}_y)$ direction through controlled site-dependent detunings $\Delta_{m,n}$, as shown in Fig.~\pref{fig3}{a} and more experimental details can be found in Supplementary Materials \cite{supp}. The resulting Hamiltonian reads
\begin{equation}
\mathcal{H}_{\rm Hall} = \mathcal{H}_{\rm HH} + \sum_{m,n} nF c^\dagger_{m,n} c_{m,n}^{}.
\end{equation} 
where $F=\partial\Delta_{m,n}/\partial n$ denotes the effective electric-field strength.

To probe the transverse response, we measure the center-of-mass (CoM) displacement along the transverse lattice direction,  $d_{\rm CoM} = \sum_{m,n} m\rho_{m,n}$. From a semiclassical perspective, this transverse drift reflects the integrated Hall velocity for the Berry curvatures sampled by the initial wavepacket at site $(0,0)$. Therefore, both the drift magnitude and direction relate to the applied synthetic electric field and the magnetic flux, and directly manifest the dynamical signature of the Hofstadter band topology. Figure \pref{fig3}{b} displays the wavepacket distributions after finite evolution times for a positive $F/h = +1~{\rm kHz}$. A clear transverse displacement along the $-\hat{e}_x$ direction is observed, consistent with the expected Hall deflection induced by the synthetic magnetic flux. Reversing the sign of the electric field leads to a corresponding reversal of the transverse drift direction, as shown in Fig.~\pref{fig3}{c}, directly demonstrating the directional nature of the Hall response.

We further investigate the dependence of the transverse transport on the effective magnetic flux $\phi$.  As summarized in Fig.~\pref{fig3}{d}, the measured time-averaged CoM drift $\langle d_{\rm CoM}\rangle_t = \frac{1}{t}\int_0^t d_{\rm Com}(\tau){\rm d}\tau$ varies approximately linearly to the flux value within the explored parameter regime, while its sign strongly depends on the direction of the synthetic electric field. The two set of measurements for $F/h = \pm 1~{\rm kHz}$ exhibit nearly perfect mirror symmetry, consistent with the expected Lorentz-type response arising from the combined action of both $F$ and $\phi$. These observations establish that the present two-dimensional MSL platform can emulate non-equilibrium transport response from non-trivial topological band structures associated with synthetic electromagnetic fields. More broadly, the ability to engineer both local gauge fluxes and controllable driving forces provides a powerful route toward studying topological transport and dynamical responses with programmable tailoring. 

~

\emph{Transport across a programmable gauge-field interface. --} 
A central capability of our platform is the realization of spatially inhomogeneous synthetic gauge fields with local, link-resolved programmability. To demonstrate this feature, we engineer a synthetic gauge-flux domain wall separating two lattice regions with opposite magnetic fluxes, as illustrated in Fig.~\pref{fig4}{a} (see Supplementary Information for the phase-programming protocol). This geometry provides a stringent test of local gauge control because the relevant observable is not the response to a global flux, but the motion of a wavepacket launched at an interface between two programmed gauge sectors. Different from the case with finite uniform flux, transport along the domain-wall boundary is now qualitatively modified by the competing opposite gauge phases of the neighboring plaquettes. In particular, the tunnelling processes parallel to the interface are phase-sensitive and would possibly avoid the strong destructive interference that suppresses bulk transport. As a result, the wavepacket dynamics is expected to become preferentially guided along the boundary, leading to anisotropic propagation confined near the interface.

\begin{figure}[]
	\centering
	\includegraphics[width=\linewidth]{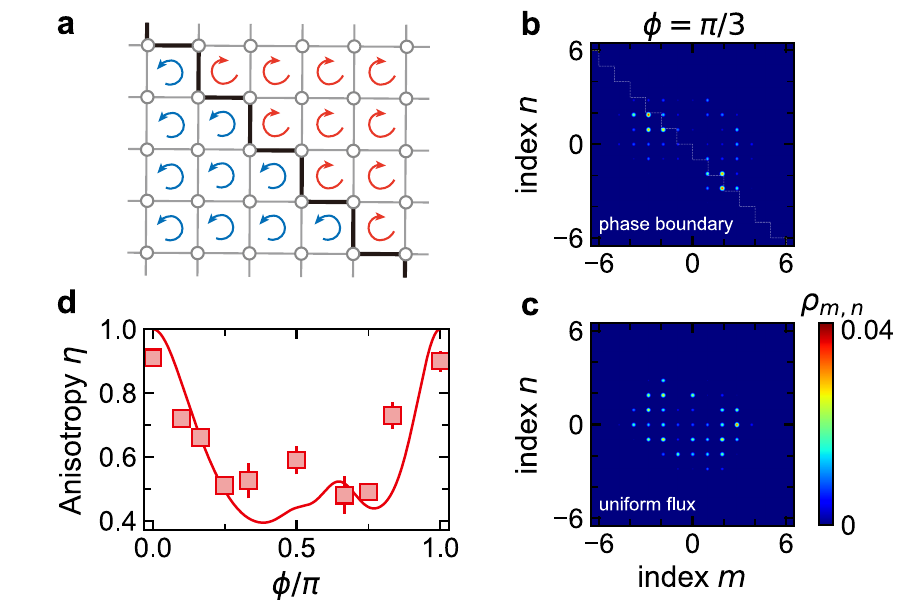}
	\caption{\textbf{Boundary-guided dynamics at an engineered gauge-field interface.} 
    \textbf{(a)} Schematic of creating a synthetic gauge-field boundary (flux interface, bold black connections) in the 2D MSL, where the synthetic magnetic flux takes opposite signs ($+\phi$ and $-\phi$) on the two sides of the diagonal interface.
    \textbf{(b)} Measured wavepacket distributions at $t=0.35~\mathrm{ms}$ for an injection at the interface center for $\phi=\pi/3$.
    \textbf{(c)} Measured wavepacket distributions for uniform $\phi=\pi/3$  without an interface as a comparison to the result in (b).
    \textbf{(d)} Anisotropy parameter $\eta$ characterizing wavepacket elongation along the interface as a function of the flux $\phi$.}
    \label{fig4}
\end{figure}

To verify this, we initialize the BEC at the center of the interface and monitor the evolution in momentum space. For $\phi=\pi/3$ , the measured wavepacket distributions at $t=0.35~\mathrm{ms}$ are shown in Figs.~\pref{fig4}{b}, while the results with uniform flux values are shown in Figs.~\pref{fig4}{c} as a comparison. With the phase domain-wall boundary, the wavepacket exhibits pronounced anisotropic expansion, propagating preferentially along the diagonal interface. The comparison with the uniform-flux evolution is essential: it shows that the anisotropy is created by the spatial gauge texture itself, rather than by the finite lattice geometry or by the flux magnitude alone. This behavior directly reflects interface-guided transport associated with edge modes localized near the sign-changing flux boundary, induced by spatially structured gauge fields.

To quantify the anisotropy of this dynamics, we explicitly define a dimensionless parameter
\begin{equation}
\eta=\frac{\sum_{m,n}\rho_{m,n}(m+n)^2}{\sum_{m,n}\rho_{m,n}(m-n)^2}.
\end{equation}
Here, $\eta=1$ corresponds to isotropic expansion, whereas $\eta<1$ indicates preferential spreading along the off-diagonal interface.
The measured values of $\eta$, summarized in Fig.~\pref{fig4}{d}, exhibit a clear suppression at intermediate flux values, confirming the increasing confinement of the dynamics to the engineered boundary. These observations demonstrate that programmable two-dimensional MSLs can realize not only uniform synthetic magnetic fields, but also sharply structured gauge-field textures supporting qualitatively distinct transport phenomena. These also identify a concrete experimental observable, boundary-guided wavepacket motion, that is directly enabled by local flux programming. More broadly, the ability to engineer spatially varying gauge configurations opens new opportunities for investigating interface physics, topological transport across designed boundaries, and emergent dynamics in synthetic systems with programmable geometry and topology.

~

\emph{Conclusion and outlook. --} 
In summary, we have realized a highly programmable 2D MSLs of ultracold $^{87}$Rb atoms and further implemented the Harper–Hofstadter model with tunable synthetic gauge fields. By engineering tunnelling amplitudes and Peierls phases on individual links, we achieved direct local control of the plaquette flux and thereby accessed both uniform and spatially structured gauge-field configurations. We have observed the crossover from ballistic expansion to strongly flux-modified bulk dynamics, demonstrated Hall-type transverse drift arising from the interplay between synthetic electric and magnetic fields, and directly revealed interface-guided transport along an engineered flux domain-wall boundary.

These results establish 2D MSLs as a versatile setup for designing and probing quantum transport under tailored, spatially structured synthetic gauge fields. Broadly speaking, the level of Hamiltonian control demonstrated here provides a route toward synthetic quantum matter with designed gauge structures beyond those naturally available in condensed-matter systems. For example, extensions to more complex lattice geometries and programmable flux distributions could enable explorations of flat-band physics in highly customized settings~\cite{Agrawal2024,Lebrat2026}. Moreover, combining the present scheme with non-Abelian gauge-field engineering in momentum space~\cite{Liang2024} may provide access to richer topological band structures and exotic transport phenomena beyond Abelian cases. Incorporating tunable atomic interactions~\cite{An2018,Xie2020,Wang2024,Chin2010} would further open the door to studies of strongly correlated topological matter, interacting quantum Hall dynamics \cite{Tai2017}, and emergent many-body phases \cite{Chen2025,Impertro2025,Kumar2016} in a clean, tunable, and dynamically reconfigurable setting.

~

\textbf{Acknowledgment. --}
We acknowledge support from the National Natural Science Foundation of China under Grants No. 92576201 and 12425408, the National Key Research and Development Program of China under Grants No. 2023YFA1406703 and No. 2022YFA1404203, the Quantum Science and Technology-National Science and Technology Major Project under Grant No. 2024ZD0300601, The Zhejiang Provincial Applied Basic Research Program under Grant No. 2026C02A2005, the Fundamental Research Funds for the Central Universities under Grant No. 2024FZZX02-01-02, and the China Postdoctoral Science Foundation under Grant No. 2024T170763 and GZB20240666. T.C. acknowledges the startup funding from Xi'an Jiaotong University and the Fundamental Research Funds for the Central Universities. 

~

\textbf{Author contributions. --}
H.W., H.L., T.C. and B.Y. conceived the project. H.W., H.L., Y.P. and Y.L. performed the experiments and analysed the data. H.W., H.L., B.G., T.C. and B.Y. contributed to the theoretical interpretation. H.W., H.L., T.C. and B.Y. wrote the manuscript with input from all authors. T.C. and B.Y. supervised the project.

~

\textbf{Competing interests. --}
The authors declare no competing interests.

~

\textbf{Data availability. --}
The data that support the findings of this study are available from the corresponding authors upon request.

~

\textbf{Code availability. --}
The code used for numerical simulations and data analysis is available from the corresponding authors upon request.

\bibliographystyle{apsrev4-2}
\bibliography{sim2dHH_mainV3}

\end{document}